\magnification=1200
\input epsf
\overfullrule=0pt
\tolerance=100000
\nopagenumbers

\font\smallbf=cmb10 at 11truept

\font\legenda=cmr10 at 11truept
\font\smalltt=cmtt10 at 11truept

\font\tenbifull=cmmib10 \skewchar\tenbifull='177
\font\tenbimed=cmmib7   \skewchar\tenbimed='177
\font\tenbismall=cmmib5  \skewchar\tenbismall='177
\textfont9=\tenbifull
\scriptfont9=\tenbimed
\scriptscriptfont9=\tenbismall

\mathchardef\alpha="710B
\mathchardef\beta="710C
\mathchardef\gamma="710D
\mathchardef\delta="710E
\mathchardef\epsilon="710F
\mathchardef\zeta="7110
\mathchardef\eta="7111
\mathchardef\theta="7112
\mathchardef\iota="7113
\mathchardef\kappa="7114
\mathchardef\lambda="7115
\mathchardef\mu="7116
\mathchardef\nu="7117
\mathchardef\micron="716F
\mathchardef\xi="7118
\mathchardef\pi="7119
\mathchardef\rho="711A
\mathchardef\sigma="711B
\mathchardef\tau="711C
\mathchardef\upsilon="711D
\mathchardef\phi="711E
\mathchardef\chi="711F
\mathchardef\psi="7120
\mathchardef\omega="7121
\mathchardef\varepsilon="7122
\mathchardef\vartheta="7123
\mathchardef\varphi="7124
\mathchardef\varrho="7125
\mathchardef\varsigma="7126
\mathchardef\varpi="7127
\def\moldura#1#2{\vbox{\hrule\hbox{\vrule\kern3pt\vbox{\kern3pt 
        \vbox{\hsize #2truecm\noindent\strut\baselineskip=8pt#1\strut} 
              \kern3pt}\kern3pt\vrule}\hrule}} 

\def\boxit#1{\vbox{\hrule\hbox{\vrule\kern3pt\vbox{\kern3pt#1\kern3pt}\kern3pt\vrule}\hrule}}

 at 8truept

\

\centerline{}
\bigskip
\bigskip
\bigskip
\baselineskip=14pt

\centerline{{\bf Dispersionless Limit of Integrable Models}}
\vfill
{\baselineskip=11pt
\centerline{J. C. Brunelli}
\medskip
\medskip
\centerline{Universidade Federal de Santa Catarina}
\centerline{Departamento de F\'\i sica -- CFM}
\centerline{Campus Universit\'ario -- Trindade}
\centerline{C.P. 476, CEP 88040-900}
\centerline{Florian\'opolis, SC -- BRAZIL}
\smallskip
\centerline{ brunelli@fsc.ufsc.br}
}
\vfill

\centerline{\bf {Abstract}}

\medskip
Nonlinear dispersionless equations arise as the dispersionless limit of well know integrable hierarchies of equations or by construction, such as the system of hydrodynamic type. Some of these equations are integrable in the Hamiltonian sense and appear in the study of topological minimal models. In the first part of the review we will give a brief introduction to integrable models, mainly its Lax representation. Then, we will introduce the dispersionless limit and show some of our results concerning the two-component hyperbolic system of equations such as the polytropic gas and Born-Infeld equations.

\medskip

\vfill
\eject

\bigskip
\noindent {\bf 1. {Introduction}}
\medskip

The study of integrable models or solvable nonlinear partial
differential equations is an active area of research since the
discovery of the inverse scattering method [1-3]. These models are in
a sense universal since they show up in many areas of physics such as
solid state, nonlinear optics, hydrodynamics, field theory just to
name a few. Also, integrable models are linked to many areas of
mathematics (see the chart in
{\smalltt http://www.ma.hw.ac.uk/solitons/procs/bullough1/bullough1/bullough1.html})
and have beautiful structures behind them.

In this review we want to approach the dispersionless limit of some
integrable models and describe some of our work on this subject
[4-7]. This review is organized as follows: In Section 2 we review or at
least introduce some basic facts on integrable models. We use the
Korteweg-de Vries equation (KdV) as an example. In Section 3 we
introduce  the dispersionless limit of an integrable model using the
KdV equation to obtain the corresponding Riemann equation. Section 4
reviews our work with a special class of dispersionless systems  known
as two-component hyperbolic systems. We show our results
concerning the Hamiltonian structures for the Riemann equation [4] the
dispersionless Lax representation for the polytropic gas dynamics
[5] and Born-Infeld equation [6]. Finally, in Section 5, we conclude
with some problems that deserve further investigations. 
\bigskip
\noindent {\bf 2. Integrable Models}
\medskip
\noindent {\bf 2.1 Solitons}
\medskip

We are interested in nonlinear partial differential equations such as
the sine-Gordon equation, nonlinear Schr\"odinger equation, 
Korteweg-de Vries equation (KdV), etc. These equations, as we will
see,  are very special since they are integrable. From now on we will
illustrate the main results concerning integrability using the KdV equation.

The KdV equation has as solution what is called today a soliton. We
can trace the discovery of the soliton back to 1834 with the Scott
Russell's  experiment [8] to generate solitary waves in water, i.e.,
localized single entity waves. A modern version of his experiment is
shown in Figure 1 (see
{\smalltt http://www.ma.hw.ac.uk/$\sim$chris/scott\_russel.html} for
an attempt to recreate Scott Russell's soliton).
\noindent Scott  Russell found that the volume $V$ of water wave is
equal to the volume of water displaced and that the speed $c$ of  the
solitary wave is related with its amplitude $a$, depth  of water $h$
and acceleration of gravity $g$ by
$$
c^2=g(h+a) \eqno(2.1)
$$
This equation shows that higher waves travel faster. Attempts to
obtain (2.1) theoretically were done by Boussinesq (1871) and Lord
Rayleigh (1876) but an equation for $u(x,t)$ in the small amplitude
($h\gg a$) and in the long wave regime ($h\ll\ell$) was deduced by
Korteweg-de Vries in 1895 [9]. This is the now famous KdV equation
$$
\moldura{$u_t=uu_x+u_{xxx}$}{3.0}\eqno(2.2)
$$
where $u(x,t)$ is the wave profile and $u_t={\partial u\over\partial
t}$, $u_x={\partial u\over\partial x}\,,\ldots\,$.
\topinsert
\epsfxsize 10.0truecm 
{\hfil\epsfbox{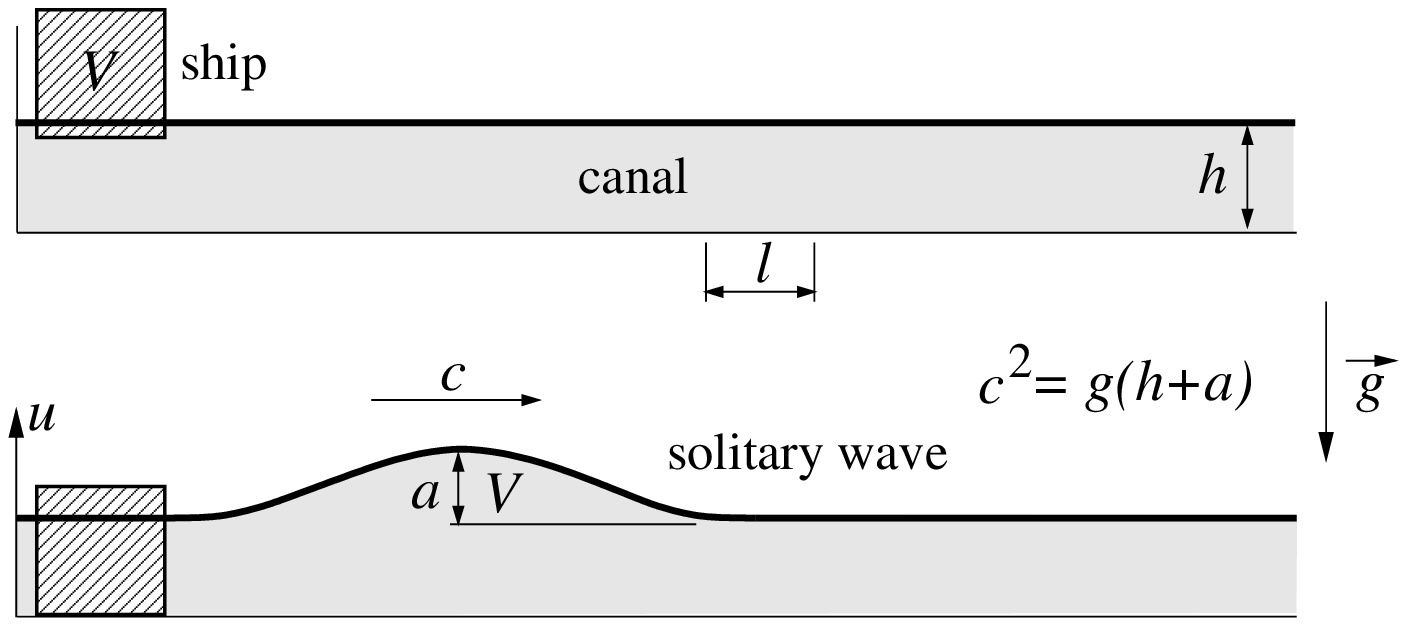}\hfil}
\vskip .5 truecm 
\noindent{\legenda {\bf Figure 1.} Generation of a solitary wave.}
\endinsert

The interest in the KdV equation $(2.2)$ was resumed after studies of 
Fermi, Pasta and Ulam in 1955 [10] on numerical models of phonons in
non-linear lattices, which are models closely related with the
discretisation of the KdV equation. Motivated by these results,
Zabusky and Kruskal in 1965 [11] studied numerically equations like
$(2.2)$ with periodic boundary conditions and were led  to introduce
the concept of ``soliton'' solutions. In 1967 Gardner, Greene, Kruskal and
Miura [12] solved equation (2.2) exactly, introducing the ``Inverse
Scattering Transform Method'' (ISTM), and were able to obtain its
analytic expression. The so called
1-soliton and 2-soliton solutions of the KdV equation (2.2), for rapidly
decreasing boundary conditions
$$
u(x,t)\to 0\quad \hbox{for}\quad x\to\pm\infty\,,\hfill
$$
are 
$$
\eqalign{
u(x,t)=&{1\over2}c^2\hbox{sech}^2\left({1\over2}c(x+c^2t)\right)\quad\to\quad{\rm
1-~soliton}\cr
\noalign{\vskip .5truecm}%
u(x,t)=&12{3+4\cosh(2x-8t)+\cosh(4x-64t)\over\{3\cosh(x-28t)+\cos(3x-36t)\}^2}
\quad\to\quad{\rm 2-~soliton}
}\eqno(2.3)
$$
\midinsert
\epsfxsize 15.0truecm 
{\hfil\epsfbox{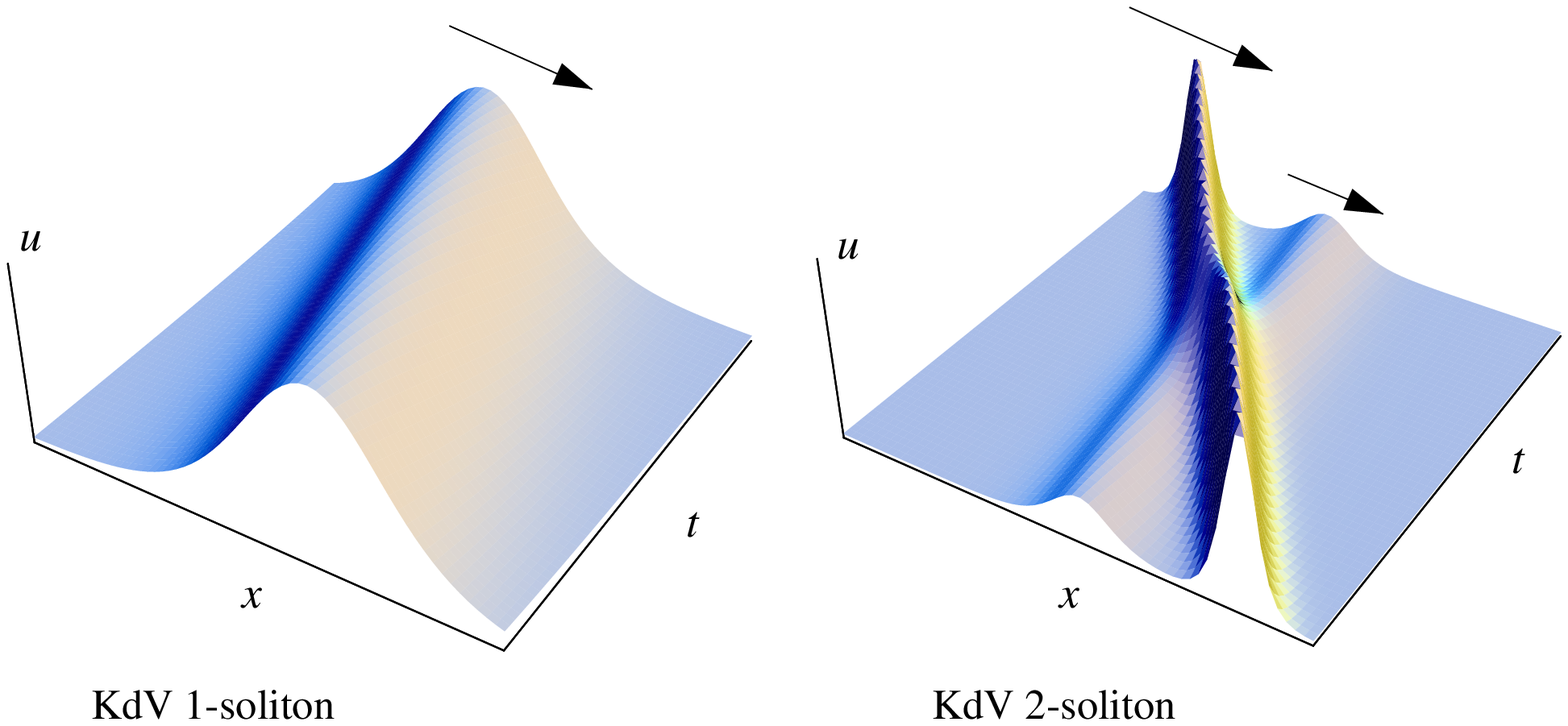}\hfil}
\noindent{\legenda {\bf Figure 2.} Time evolution for the solitons
of the KdV equation.}
\endinsert
In Figure 2 we have pictures for the time evolution of the KdV
solitons (2.3) (for some brief solitons movies see
{\smalltt http://www.ma.hw.ac.uk/solitons} and
{\smalltt
http://www.physics.otago.ac.nz/Physics100/simulations/Gamelan/java/toda}).
The 1-soliton solution in Figure 2 is the solitary wave obtained in
the Scott Russell's experiment. Observe that as the time evolves the
wave keeps its form. For the 2-soliton solution in Figure 2, since the
taller the soliton the faster it moves, the two solitons will interact
nonlinearly when they meet. But, the amazing fact is that the two
solitons will almost keep their initial form after interaction, there
will be only a shift in their positions. This particle-like
character and ability to retain its identity after interactions is what
characterize a soliton solution of a nonlinear equation such as the
KdV one.
\bigskip
\noindent {\bf 2.2 Inverse Scattering}
\medskip

The next breakthrough in the soliton thread came in 1968 with the Lax [13]
discovery about the meaning of the ISTM. His observation is that the
KdV equation has the representation
$$
\moldura{$\displaystyle{\partial L\over\partial t}=[B,L]$}{2.4}
\eqno(2.4)
$$
where
$$
\eqalign{
 L=&\partial^2+{1\over 6}u\cr
B=&4\,\partial^3+{1\over2}(\partial u+u\partial)\cr
}\eqno(2.5)
$$
are operators. Here $\partial\equiv{\partial\ \over\partial x}$
satisfies $\partial f=f_x+f\partial$. We call $L$ the Lax operator
and in some sense we can find a Lax representation such as $(2.4)$ for any
integrable system. In this way, starting from $(2.4)$, we can apply the
ISTM for other nonlinear equations.

We can write the following eigenvalue problem for the Lax operator $L$
$$
L\psi=-\lambda\psi\eqno(2.6)
$$
It is easy to see that since $L$ evolves in time as $(2.4)$ we have
$\lambda_t=0$, i.e., the eigenvalue problem is isospectral. For the
KdV equation (2.6) assumes the form
$$
{\partial^2\psi\over\partial x^2
}+\left({1\over6}u(x,t)+\lambda\right)\psi=0\eqno(2.7)
$$ 
which is the time-independent Schr\"odinger equation and where $t$ is a
parameter (not the time in the Schr\"odinger equation). Now we can
obtain a solution $u(x,t)$ as follows: For some given initial
condition $u(x,0)$ we solve $(2.7)$ and obtain the scattering data
$S(t=0)$, since $u$ satisfies the KdV equation we can obtain the
scattering data for any $t$, so from $S(t)$ we use the inverse
scattering (as we usually do in quantum mechanics) to find the
``potential'' $u(x,t)$ from the scattering data $S(t)$. This is the
ISTM routine and the main steps are illustrated in the diagram bellow.
\medskip
\centerline{\moldura{$$
\matrix{
\hbox{initial condition}& &\hbox{scattering data can}&\cr
\hbox{(given)}&&\hbox{be calculated}&\cr
\Uparrow & & \Uparrow&\cr\cr
u(x,0) &\longrightarrow &S(t=0)&\cr\cr
\uparrow& &\downarrow&\Rightarrow\hbox{KdV}\hfill\cr\cr
\moldura{$u(x,t)$}{1.1} &\longleftarrow & S(t)&\cr
&\hbox{inverse}&&\cr
&\hbox{scattering}&&}
$$}{11.5}}
\centerline{\bf Inverse Scattering Transform Method}
\bigskip
\noindent {\bf 2.3 Hamiltonian Systems}
\medskip

In 1970 Gardner [14] showed that the KdV equation is a Hamiltonian
integrable system. Then, Faddeev and Zakharov in 1971 [15] were able to
interpret the ISTM as a change of variables to the action angle
variables. In fact, the representation of integrable models as
integrable Hamiltonian systems is the starting point to the ``Quantum
Inverse Scattering Method''.
Before we see how the KdV equation can be expressed in Hamiltonian form
let us review the symplectic formalism for Hamiltonian systems. A
Hamiltonian system is described by a phase space $q_i,p_i$, with
$i=1,\dots,N$, and a Hamiltonian function $H(p_i,q_i)$. The equations of
motion are then given by the Hamilton's equations
$$
\eqalign{
{\dot  q}_i&={\partial H\over\partial p_i}\cr
{\dot  p}_i&=-{\partial H\over\partial q_i} 
}\eqno(2.8)
$$
Alternatively, we can describe a Hamiltonian system using Poisson
brackets, for the dynamical variables $A(q,p)$ and $B(q,p)$, defined by
$$
\{A,B\}={\partial A\over\partial q_i}
{\partial B\over\partial p_i}-{\partial A\over\partial p_i}
{\partial B\over\partial q_i}\eqno(2.9)
$$
which is skew-symmetric and satisfies the Jacobi identity. The
variables of phase space satisfy the canonical relations
$\{q_i,q_j\}=\{p_i,p_j\}=0$ and 
$\{q_i,p_j\}=\delta_{ij}$. The Hamilton's equations $(2.8)$ assume the form
$$
\eqalign{
{\dot  q}_i=&\{q_i,H\}\cr
{\dot  p}_i=&\{p_i,H\}
}
\eqno(2.10)
$$
Putting the variables $q_i$ and $p_i$ in an $2N$ dimension
column $z$ the equations $(2.8)$ assume  the form
$$
{d\ \over dt}
\underbrace{\left(\matrix{q_1\cr\vdots\cr q_N\cr p_1\cr\vdots\cr p_N}\right)}_{\displaystyle\equiv z}=\underbrace{\left(\matrix{0 & I\cr -I & 0}\right)}_{\displaystyle\equiv J}\underbrace{\left(\matrix{{\partial/\partial q_1}\cr\vdots\cr{\partial/\partial q_N}\cr {\partial/\partial p_1}\cr\vdots\cr {\partial/\partial p_N}}\right)}_{\displaystyle\equiv{\vec \nabla}}H\eqno(2.11)
$$
or
$$
{\dot z}^a=J^{ab}\partial_b H\qquad a,b=1,\dots,2N\eqno(2.12)
$$
and even in a more compact form as
$$
\moldura{${\dot z}=J{\vec \nabla}H$}{1.8}\eqno(2.13)
$$
This is the symplectic formalism for Hamiltonian systems. The Poisson
brackets can be written as
$$
\{A,B\}=\left({\vec\nabla}A\right)^t J\left({\vec\nabla}B\right)\eqno(2.14)
$$
where $J^{ab}=-J^{ba}$ and 
$\sum\left(J^{ab}\partial_d J^{bc}+{\rm cyclic}\right)=0$. The canonical
relations are given by $\{z,z\}=J$ and $(2.10)$ by
$$
\moldura{${\dot z}=\{z,H\}$}{2.0}\eqno(2.15)
$$

We can perform some generalizations, allowing $J$ to depend on $z$,
$J(z)$, and going from a discret sympletic space, of dimension $2N$, to
the continuum where we have now a field $u(x,t)$ instead of
$z(t)$. Then, we have the following ``dictionary''
\vskip .3 truecm
\centerline{\moldura{$$
\matrix{
z(t) & \rightarrow & u(x,t)\cr\cr
H(z) & \rightarrow & H[u]\hbox{ functional}\cr\cr
{\vec \nabla} H & \rightarrow &\displaystyle 
{\delta H\over \delta u}\hbox{ functional derivative}\cr\cr
J(z)\hbox{ skew-symmetric matrix}&\rightarrow & {\cal D}(u)\hbox{
skew-adjoint operator}\cr\cr
{\dot z}=J{\vec \nabla}H&\rightarrow &\displaystyle{\dot u}={\cal D}{\delta H[u]\over \delta u}\cr\cr
\{z,z\}=J(z)&\rightarrow &\displaystyle\{u(x),u(x')\}={\cal D}
\delta(x-x')\cr\cr
\{A,B\}=\left({\vec\nabla}A\right)^t J\left({\vec\nabla}B\right)&\rightarrow & 
\displaystyle\{A[u],B[u]\}=\int dx
{\delta A\over \delta u}{\cal D}{\delta B\over \delta u}\cr\cr
}
$$}{14}}
\vskip .5truecm
\noindent If there is a $J^{-1}$ we say that we are in a symplectic manifold,
otherwise we are in a more general situation of a Poisson
manifold. Note that the functional derivative ${\delta H[u]\over
\delta u}$ is defined as
$$
{\delta H[u(x)]\over \delta u(y)}=\lim_{\epsilon\to 0}{H[u(x)+\epsilon\delta(x-y)]-H[u(x)]\over\epsilon}\eqno(2.16)
$$
which for $H[u]=u(x)$ yields
$$
{\delta H[u(x)]\over \delta u(y)}=\delta(x-y)
$$
and for $H[u]=\int dx\, h(x,u,u_x,u_{xx},\dots)$
$$
{\delta H[u(x)]\over \delta u(y)}=\left({\partial\ \over\partial u}-{\partial\ \over\partial x}{\partial\ \over\partial u_x}+{\partial^2\ \over\partial x^2}{\partial^2\ \over\partial u_{xx}^2}+\dots\right)h 
$$
where the right hand side is just the Euler-Lagrange operator acting
on $h$.

Now, let us return to the KdV equation $(2.2)$ and observe that it can
be rewritten as
$$
\eqalign{
u_t=&uu_x+u_{xxx}\cr
=&{\partial\ \over\partial x}\left({1\over2}u^2+u_{xx}\right)\cr
}\eqno(2.17)
$$
Introducing the Hamiltonian
$$
H_2=\int dx\left({1\over3!}u^3-{1\over2}u_x^2\right)\eqno(2.18)
$$
we see that ${\delta H_2\over\delta u}={1\over2}u^2+u_{xx}$ and
${dH_2\over dt}=0$. The operator
$$
{\cal D}_1={\partial\ \over\partial x}
\eqno(2.19)
$$ 
is skew-adjoint and satisfies the Jacobi identity. So, (2.17) can be
written in Hamiltonian form as
$$
u_t={\cal D}_1{\delta H_2\over\delta u}=\{u(x),H_2\}_1\eqno(2.20)
$$
where
$$
\{u(x),u(y)\}_1={\cal D}_1\delta(x-y)\eqno(2.21)
$$
and we are omitting the explicit dependence on $t$.

Besides $(2.18)$ the KdV equation $(2.2)$ has an infinite number of
conserved charges
$$
\eqalign{
H_0=&\int dx\, u\cr
H_1=&\int dx\, {1\over 2}u^2\cr
H_2=&\int dx\,\left({1\over 3!}u^3-{1\over 2}u_x^2\right) \cr
H_3=&\int dx\,\left({1\over 4}u^4-3uu_x+{9\over 5}u^2_{xx}\right) \cr
H_4=&\int dx\, \left({1\over 5}u^5-6u^2u_x^2+{36\over 5}uu^2_{xx}-
{108\over 35}u^2_{xxx}\right)\cr
\vdots&
}\eqno(2.22)
$$
and it can be shown that these charges are in involution, i.e.,
$$
\{H_n,H_m\}_1=0\eqno(2.23)
$$
making the KdV equation integrable in Lioville's sense.

In 1978 Magri [16] discovered that equations like KdV have a second
Hamiltonian structure. The operator
$$
{\cal D}_2={\partial^3\ \over\partial x^3}+{1\over3}\left({\partial\
\over\partial x}u+u{\partial\ \over\partial x}\right)\eqno(2.24)
$$
is skew-adjoint and satisfies Jacobi identity, and the KdV equation can
be written in the alternative Hamiltonian form
$$
u_t={\cal D}_2{\delta H_1\over\delta u}=\{u(x),H_1\}_2\eqno(2.25)
$$
where
$$
\{u(x),u(y)\}_2={\cal D}_2\delta(x-y)\eqno(2.26)
$$
These charges $(2.22)$ are also in involution with respect to this
second Hamiltonian structure
$$
\{H_n,H_m\}_2=0\eqno(2.27)
$$
We say that the KdV equation is a bi-Hamiltonian system. In general we
say that a system is bi-hamiltonian if there are Hamiltonian operators
${\cal D}_1$ and  ${\cal D}_2$  which are compatible, i.e., such that
${\cal D}_1$, ${\cal D}_2$ and  $\lambda_1{\cal D}_1+\lambda_2{\cal
D}_2$  satisfy the Jacobi identity. It can be shown [16] that if a system
is bi-Hamiltonian it is integrable in Lioville's sense.

Starting with the works of Gel'fand and Dickey in 1975 [17], Adler in
1979 [18]
and many others, algebraic developments started to take place. The key
role played by the Lax operator $L$, in obtaining the conserved charges
$H_n$, the Hamiltonian  structures, the hierarchy of equations that
share $H_n$ was then revealed. In the next sections we will introduce and
apply some of these techniques in the dispersionless situation.

\bigskip
\noindent {\bf 3. Dispersionless Limit}
\medskip

We have seen that solitons preserve their shape and speed after
collision. The soliton solution has a nondispersive nature. This is so
not because dispersion effects are absent but because there is a
compensation by the nonlinearities of the system. Let us look at the
KdV equation (2.2) more closelly. If we eliminate the nonlinear term
in (2.2) we get the linear dispersive equation
$$
u_t=u_{xxx}\eqno(3.1)
$$
which admits the solution
$$
u(x,t)=\!\!\int\!\! dk\,A(k)\hbox{e}^{i(kx-w(k)t)}\eqno(3.2)
$$
This is a pure dispersive solution. In Figure 3 we see that a initial
configuration at $t=0$ will disperse as time goes on. Eliminating the
dispersive term we get the pure nonlinear equation
$$
u_t=uu_x\eqno(3.3)
$$
It can be easily checked by substitution that
$$
u(x,t)=f(x-ut)\eqno(3.4)
$$
with $f$ arbitrary, satisfies $(3.3)$. From this solution we conclude that
the velocity of a point of the wave, with constant amplitude $u$, is
proportional to its amplitude leading to the ``breaking'' of the wave,
as shown in Figure 3. The wave also  develops discontinuities
(indicated by the vertical dashed line in Figure 3) in its
evolution. The ``miracle'' of the soliton solution is due to a balance
between the dispersion and the breaking of the wave, both phenonema
placed together lead to the wave profile
to propagate without changing its shape.
\midinsert
\epsfxsize 14.5truecm 
{\hfil\epsfbox{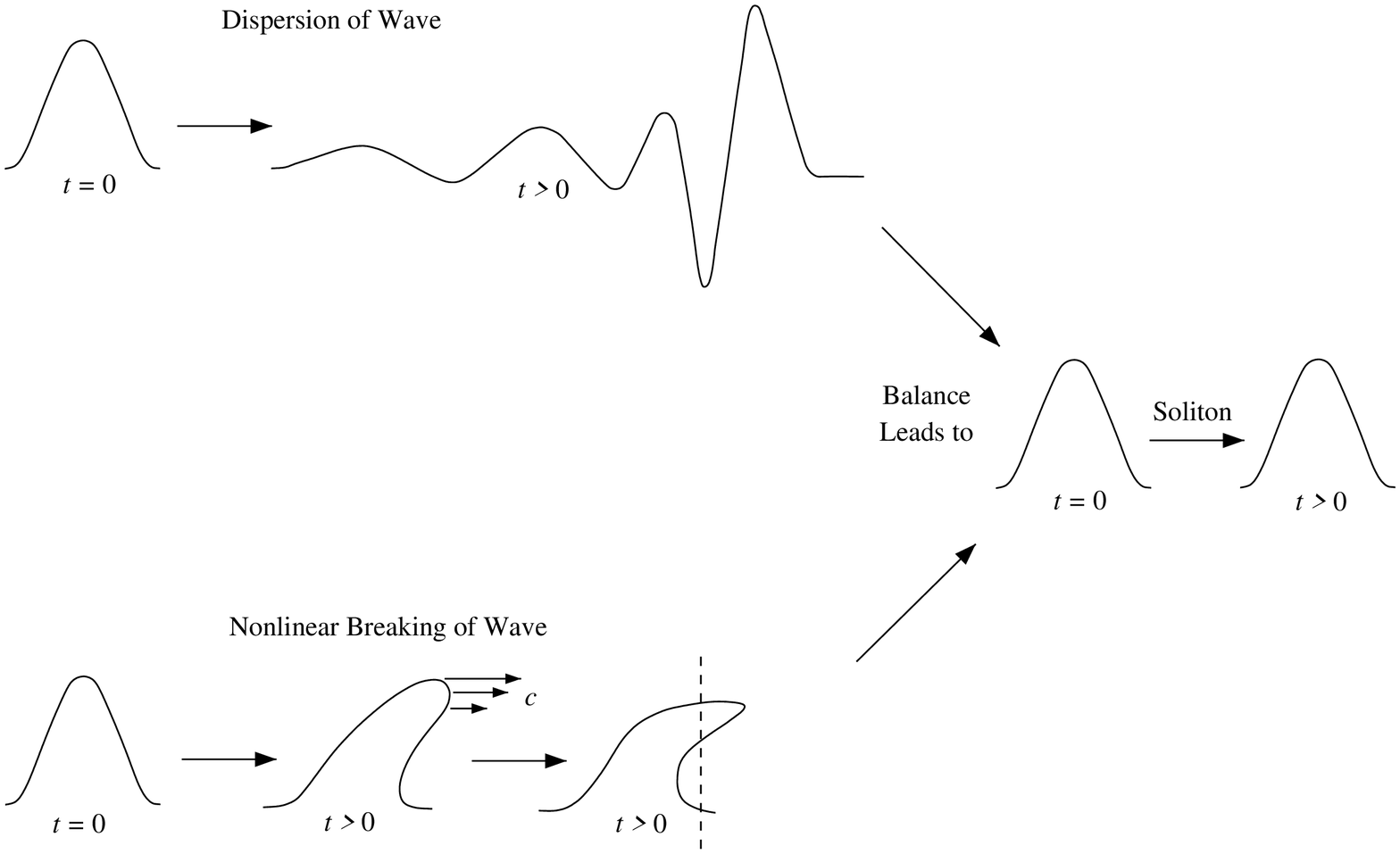}\hfil}
\vskip .5 truecm 
\noindent{\legenda {\bf Figure 3.} The balance effects of dispersion
and breaking in a soliton.}
\endinsert

Equation (3.3) is called the dispersionless KdV or Riemann
equation [19]. The interesting fact is that this equation is a
integrable Hamiltonian system. We will return to study this equation
in the next section but for the moment let us analyse how we get
dispersionless equations. Dispersionless equations can be obtained by
construction or as a quasi-classical limit of integrable ones [20]
. In the latter case we make the scaling
${\partial\ \over\partial t}\to\alpha{\partial\ \over\partial t}$, 
${\partial\ \over\partial x}\to\alpha{\partial\ \over\partial x}$    
and take the limit $\alpha\to 0$. For the KdV equation $(2.2)$ (we will
change the constant factors on it, so instead of (2.5) we have
$L=\partial^2+u$
and $B=\partial^3+{3\over4}(\partial u+u\partial))$
$$
\matrix{
4u_t=u_{xxx}+6uu_x&\Rightarrow&4\alpha u_t=\alpha^3 \hbox{\rlap{$\nearrow^0$}{$u_{xxx}$}}+6\alpha u_x u\cr\cr
\Downarrow&&u_t={3\over2}uu_x\cr\cr
\rm KdV&&\Downarrow\cr\cr
&&\rm Riemann
}
$$
This is like the WKB aproximation in quantum mechanics and we will use
it as our guideline [20].

Dispersionless integrable systems were introduced by Lebedev and Manin
[21] and Zakharov [22], and although interesting on their own started
to appear recently in developments in low-dimensional quantum field
theory. It has been shown that there is a connection
between 2-dimensional field theories and integrable equations of
hydrodynamical type [23-25] (which are dispersionless systems). In
2-dimensional topological field theories [26]  we are interested in
calculating, from the partition function
$$
Z_M=\int [d\phi]\,\hbox{e}^{-S[\phi]}\eqno(3.5)
$$
the correlation functions
$$
\langle\phi_\alpha(x)\phi_\beta(y)\cdots\rangle_M=\langle\phi_\alpha\phi_\beta\cdots\rangle_M\eqno(3.6)
$$
which depend only on the topology of the manifold $M$. The 2-point and 3-point
correlation functions are given respectively by [27]
$$
\eqalign{
\langle\phi_\alpha\phi_\beta\rangle&=\eta_{\alpha\beta}=\hbox{nondegenerate constant}\cr
\langle\phi_\alpha\phi_\beta\phi_\gamma \rangle&=c_{\alpha\beta\gamma}(t)=
{\partial^3 F(t)\over\partial t^\alpha \partial t^\beta \partial t^\gamma }
\qquad\alpha,\beta,\gamma=1,2,\dots,n
}\eqno(3.7)
$$
where $t=(t^1,t^2,\dots,t^n)$ are the coupling constants and $F(t)$
 is the free energy. The correlations (3.7) define a commutative and
 associative algebra (with an identity)
$$
e_\alpha\circ e_\beta=c^\gamma_{\alpha\beta}e_\gamma\eqno(3.8)
$$
with $e_\alpha$ defining a basis for the algebra. The associativity of
the algebra, $(e_\alpha\circ e_\beta)\circ e_\gamma=e_\alpha\circ(
e_\beta\circ e_\gamma)$, gives
$$
{\partial^3 F(t)\over \partial t^\alpha \partial t^\beta \partial t^\lambda}
\,\eta^{\lambda\mu}
{\partial^3 F(t)\over \partial t^\gamma \partial t^\delta \partial t^\mu}=
{\partial^3 F(t)\over \partial t^\gamma \partial t^\beta \partial t^\lambda}
\,\eta^{\lambda\mu}
{\partial^3 F(t)\over \partial t^\alpha \partial t^\delta \partial t^\mu}
\eqno(3.9)
$$
These are the Witten-Dijkgraaf-Verlinde-Verlinde (WDVV) equations
[26,27] and can be identified with equations of hydrodynamic
type. So, solutions of hydrodynamic equation can be identified with
particular solutions of the topological field theory [25].

\bigskip
\noindent {\bf 4. Two-Component Hyperbolic Systems}
\bigskip

In a series of papers [28-31] Nutku and collaborators started to study
dispersionless systems of equations that are in-between the simple Riemann equation
[19] and the more general equations of hydrodynamic type [24]

\centerline{\moldura{\moldura{\moldura{Riemann}{1.6}\break Two Component Hyperbolic}{5.5}\break Equations of Hydrodynamic Type\break\smallskip
$u^i_t=V^{ij}(u)u^j_x\quad i,j=1,\dots,n$}{7.0}}
\medskip
\noindent In Figure 4 we can find a chart with the main
equations, of the two-component hyperbolic system type, studied
on these papers.
\pageinsert
$$
\moldura{$$
\matrix{
&\scriptstyle\hbox{\smallbf Two-Component Hyperbolic}\cr
&
\moldura{$\eqalign{u_t=& H_{uv}u_x+H_{vv}v_x\cr
v_t=& H_{uu}u_x+H_{uv}v_x\cr}$}{3.8}^{\displaystyle
*}&\eqalign{\rightarrow &\quad u_t={3\over 2}uu_x\,^\dagger}\cr
\noalign{\vskip 3pt}%
&&\scriptstyle\rm \ \ \ Riemann\cr
\qquad\qquad\swarrow&\big\downarrow&\searrow\qquad\qquad\cr
\scriptstyle H(u,v)={u\over v}+{v\over u}&\eqalign{\scriptstyle H(u,v)=&\scriptstyle-\left({1\over2}u^2 v+F(v)\right)\cr \scriptstyle f(v)=&\scriptstyle F''(v)}&
\eqalign{\scriptstyle H(u,v)=&\scriptstyle {1\over 2}u^2+{\widetilde F}(v)\cr\scriptstyle \sigma(v)=&\scriptstyle {\widetilde F}'(v)}\cr\cr
\downarrow&\big\downarrow&\big\downarrow\cr
\moldura{$\eqalign{u_t=& \left({1\over u^2}+{1\over v^2}\right)u_x-{2u\over v^3}v_x\cr
v_t=&\left({1\over v^2}+{1\over u^2}\right)v_x -{2v\over u^3}u_x\cr}$}{5.3}^{\ddagger}&
\moldura{$\eqalign{u_t+uu_x+f(v)v_x=&0\cr
v_t+(uv)_x=&0\cr}$}{4.2}&\moldura{$\eqalign{u_t=&[\sigma(v)]_x\cr v_t=&u_x}$}{2.3}\cr
\scriptstyle\rm Born-Infeld &\scriptstyle\rm  Gas\ Dynamics &\scriptstyle\rm Elastic\ Media\cr
\big\downarrow&\big\downarrow &\big\downarrow\cr
\eqalign{\scriptstyle{\widetilde u}=&\scriptstyle-(u^{-2}+v^{-2})\cr\scriptstyle{\widetilde v}=&\scriptstyle{1\over2}uv }&\scriptstyle F(v)={v^\gamma\over\gamma(\gamma-1)}\quad\gamma\ge2&\scriptstyle {\widetilde F}(v)={v^{\gamma-1}\over(\gamma-1)(\gamma-2)}\cr
 \scriptstyle\rm Verosky\ Transformation& &\cr
\big\downarrow&\big\downarrow &\big\downarrow\cr 
\eqalign{\ \ \ \qquad\gamma&=-1\qquad{\longleftarrow}\cr
\ \ {\scriptstyle\rm Chap} &  {\scriptstyle\rm lygin\ Gas}\cr\cr\cr }
&\moldura{$\eqalign{u_t+uu_x+v^{\gamma-2}v_x=&0\cr
v_t+(uv)_x=&0\cr}$}{4.3}^\dagger&\moldura{$\eqalign{u_t=&v^{\gamma-3}v_x\cr
v_t=&u_x}$}{2.2}^\dagger\cr
\noalign{\vskip -.8truecm}
&\scriptstyle\rm  Polytropic\ Gas &\scriptstyle\rm  Polytropic\ Elastic\cr
\noalign{\vskip -.2truecm}
& &\scriptstyle\rm Media\cr
&\downarrow&\cr
&\gamma=2&\cr
& \scriptstyle\rm Shallow\ Water &\cr
\noalign{\vskip -.2truecm}
&\scriptstyle\rm (Dispersionless\ TB)&
}
$$
\leftline{\quad$*\ {\rm \scriptstyle bi-Hamiltonian\ if\ } {H_{uu}\over H_{vv}}={\lambda(u)\over \mu(v)}$}
\leftline{\quad$\dagger\ {\rm \scriptstyle quadri-Hamiltonian}$}
\leftline{\quad$\ddagger\ {\rm \scriptstyle six-Hamiltonian}$}
}{15.5}
$$
\vskip -2.5 truecm
\noindent{\quad\legenda {\bf Figure 4.} Two-component hyperbolic equations.}
\endinsert

A wealth of results concerning the integrability of these systems were
revealled. Infinitely many conservation laws and multi-Hamiltonian
structures were obtained. In this section we will be interested in
reproduce some of these results from an algebraic point of view. In
order to achive this goal we must understand the Lax representation
for these systems.
\bigskip
\noindent {\bf 4.1 Riemann Equation}
\medskip

The Riemann equation
$$
\moldura{$u_t={3\over 2}uu_x$}{2.0}\eqno(4.1)
$$
is the prototype for the hyperbolic systems. We address the following
question: Is there a Lax representation for $(4.1)$? Yes, and we can
obtain it performing the semiclassical limit [20] explained in Section
3. So, if the KdV equation goes to the Riemann equation (4.1) in the
semiclassical limit, the Lax operator $L=\partial^2+u$ and
$B=\partial^3+{3\over4}(\partial u+u\partial)$ goes to the polynomials
in the variable $p$
$$
\eqalign{
E&=p^2+u\cr
M&=p^3+{3\over2}up}
\eqno(4.2)
$$
and the Lax representation (2.4) goes to
$$
{\partial E\over \partial t}=\{M,E\}\eqno(4.3)
$$
called dispersionless Lax representation (note the resemblance when we
pass from quantum to classical mechanics doing $\partial\to p$ and $[\ ,\ ]\to\{\ ,\, \}$). Here
$$
\vbox{\moldura{
$\{A(x,p),B(x,p)\}=\displaystyle{\partial A\over\partial p}{\partial B\over\partial x}-
{\partial B\over\partial p}{\partial A\over\partial x}
$}{7.1}
}\eqno(4.4)
$$
is the dispersionless Poisson bracket [21,22]. So, if we substitute
$(4.2)$ in $(4.3)$ we get $(4.1)$.

From now on we will apply some of the techniques described in [17] and [18] in
a very informal way, since we want to give only a flavor of
how the ``machinery'' works.

Let us calculate the square root of $E$ in $(4.2)$. So, we write the Laurent
polynomial 
$$
E^{1/2}=p+a_0+a_1p^{-1}+a_2p^{-2}+a_3p^{-3}+\cdots\eqno(4.5)
$$
and from $E=E^{1/2}E^{1/2}$ we obtain $a_0,a_1,a_2,a_3,\dots$, or
equivalently,  we perform a series expansion for $p\to\infty$
$$
E^{1/2}=\left[p^2\left(1+{u\over
p^2}\right)\right]^{1/2}{\buildrel p\to\infty\over=}p+{1\over 2}up^{-1}-{1\over8}u^2p^{-3}+{1\over 16}u^3 p^{-5}-{5\over 128}u^4 p^{-7}+\cdots\eqno(4.6)
$$
Now we calculate $E^{3/2}=E^{1/2}E$, $E^{5/2}=E^{1/2}E^2$ and so on
$$
\eqalign{
E^{3/2}=&p^3+{3\over 2}up+{3\over8}u^2p^{-1}+\dots\cr
E^{5/2}=&p^5+{5\over 2}up^{3}+{15\over8}u^2p+{5\over 16}u^3 p^{-1}+\cdots\cr
\vdots&
}\eqno(4.7)
$$
The set of general Laurent polynomial
$A=\sum_{i=-\infty}^{+\infty}a_ip^i$ gives rise to an associative
algebra $g=\{A\}$. This algebra can be written as a direct sum
$g=g_+\oplus g_-$, where $g_+=\{A_+\}$ and $g_-=\{A_-\}$ with
$A_+=\sum_{i\ge0}a_ip^i$ and $A_-=\sum_{i<0}a_ip^i$, respectively. We
can recognize $M$ in $(4.2)$ as
$$
M=(E^{3/2})_+\eqno(4.8)
$$
In fact, from $(4.3)$ we are motivated to write
$$
\matrix{
\displaystyle{\partial E\over\partial t}=\{(E^{1/2})_+,E\}&\Rightarrow &u_t=u_x\cr\cr
\displaystyle{\partial E\over\partial t}=\{(E^{3/2})_+,E\}&\Rightarrow &\displaystyle u_t={3\over 2}uu_x\cr\cr
\displaystyle{\partial E\over\partial t}=\{(E^{5/2})_+,E\}&\Rightarrow &\displaystyle u_t={15\over 8}u^2u_x\cr\cr
&\vdots\cr\cr}\eqno(4.9)
$$
and we have  a hierarchy of equations. We call it dispersionless KdV
(or  Riemann) hierarchy and we
write
$$
{\partial E\over\partial t_k}=\{(E^{{2k+1}\over2})_+,E\}\,,\quad
k=0,1,2,3,\ldots\eqno(4.10)
$$
treating $u$ as a function of $k+1$ variables
$$
u=u(x,t_0,t_1,t_2,\dots)
\eqno(4.11)
$$
For each $t_k$ we have what is called a flow and it can be shown that
they commute
$$
{\partial^2E\over\partial t_\ell\partial t_k}=
{\partial^2E\over\partial t_k\partial t_\ell}\eqno(4.12)
$$
consequently,  the whole set of equations $(4.9)$ is integrable since,
as we  have already pointed out, the Riemann equation is an integrable
Hamiltonian system (all the equations in $(4.9)$ share the same set of
conserved charges).

The Riemann equation can be put in the form
$$
u_t={3\over 2}uu_x={3\over 4}(u^2)_x\eqno(4.13)
$$
It follows that the quantity  $H\propto\int dx\,u^2$ is conserved. In fact
$\int dx\,u^n$ are conserved as we can show explicitly. 
These conserved charges can also be obtained from $E$. Let be $A$ any general
Laurent polynomial
$$
A=\cdots+a_{-1} p^{-1}+\cdots\eqno(4.14)
$$
following [18] we introduce the Adler's trace as
$$
\hbox{Tr}A=\int dx\,\hbox{Res}A=\int dx\, a_{-1}\eqno(4.15)
$$
which satisfies the usual relation $\hbox{Tr}{AB}=\hbox{Tr}{BA}$. From
$(4.6)$ and $(4.7)$ we see that
$$
\eqalign{
\hbox{Tr}E^{1/2}=&{1\over2}\int dx\,u\cr
\hbox{Tr}E^{3/2}=&{3\over8}\int dx\,u^2\cr
\hbox{Tr}E^{5/2}=&{5\over16}\int dx\,u^3\cr
\vdots&\cr
}
$$
and we have
$$H_n={2\over n}\underbrace{\hbox{Tr}E^{n/2}}_{p\to\infty}
\eqno(4.16)
$$
whit $\dot{H}_n=0$. From a Hamiltonian point of view the Riemann
equation is a quadri-Hamiltonian system [30]. There are Hamiltonian
operators ${\cal D}_1$,  ${\cal D}_2$, ${\cal D}_3$ which are
compatible and another Hamiltonian operator ${\cal E}$ which is
compatible only with ${\cal D}_1$. We can write
$$
\moldura{$\displaystyle u_t={\cal D}_1{\delta H_5\over\delta u}={\cal D}_2{\delta H_3\over\delta u}={3\over 4}{\cal D}_3{\delta H_1\over\delta u}={35\over 8}{\cal E}{\delta H_9\over\delta u}$}{8.8}\eqno(4.17)
$$
where
$$
\matrix{
H_1=\int dx\,u\,,\hfill &{\cal D}_1=2\partial\hfill\cr
\displaystyle H_3={1\over4}\int dx\,u^2\,,\hfill &{\cal
D}_2=u\partial+\partial u\hfill\cr
\displaystyle H_5={1\over8}\int dx\,u^3\,,\hfill &{\cal D}_3=u^2\partial+\partial u^2\hfill\cr
\displaystyle H_9={7\over128}\int dx\,u^5\,,\hfill &{\cal E}=\partial{1\over u_x}
\partial{1\over u_x}\partial\hfill\cr
}\eqno(4.18)
$$
Hamiltonian structures can also be obtained from the Lax operator $L$
($E$ in the dispersionless case). They are the symplectic structures of
Kostant-Kirillov [32] on the orbits of the coadjoint representation of
Lie groups [18,33]. For dispersionless equations the corresponding Lie
algebra is given by the associative algebra of Laurent polynomials
endowed with the bracket $(4.4)$. For the KdV equation the Lie algebra
is given by the algebra of the pseudo-differential operators with the
usual commutator. Following this scheme the Hamiltonian structures
${\cal D}_1$,  ${\cal D}_2$, ${\cal D}_3$ 
can be derived (see [4] for details) while  we were not able
to  obtain ${\cal E}$ from this scheme.

\bigskip
\noindent {\bf 4.2 Polytropic Gas Equation}
\medskip

We will try to apply the results of the last Section to some others
dispersionless equations, such as the ones in the chart of Figure
4. The polytropic gas dynamics equation
$$
\vbox{\moldura{$\eqalign{
u_t+uu_x+v^{\gamma-2}v_x=&0\cr
v_t+(uv)_x=&0
}$}{4.3}}\quad\gamma\ge2\eqno(4.19)
$$
was studied from a Hamiltonian point of view in [30]. In $(4.19)$ $u$ is
the velocity of the fluid, $v$ is its density, $f=v^{\gamma-2}$ and is
related to the pressure ($f(v)={p'(v)\over v}$) and $\gamma$ is the
ratio of specific heats (we call an ideal gas polytropic if the
specific heats are constant over a large range of temperature).

The first step will be to derive a Lax representation for $(4.19)$. We
get a hint if we consider $\gamma=2$ in $(4.19)$. In this case we
have the shallow water equation [19] also known as the
irrotational Benney equation [34]. Even though we do not know the
dispersive system which originates $(4.19)$ for any $\gamma$ we do
know it for the case $\gamma=2$. This is the dispersive shallow water
[35] equation, also called the two boson equation in field theory
$$
\eqalign{{\partial J_0\over\partial t}=&(2J_1+J_0^2+J_0')'\cr
{\partial J_1\over\partial t}=&(2J_0J_1+J_1')'}
\eqno(4.20)
$$
\vfill\eject
\noindent This equation has the following nonstandard Lax representation [36,37]
$$
\eqalign{
L=&\partial-J_0+\partial^{-1}J_1\cr
{\partial L\over\partial t}=&[L,(L^2)_{\ge1}]
}\eqno(4.21)
$$
where $(L^2)_{\ge1}$ stands for the purely nonnegative (without $p^0$
terms) part of the polynomial in $p$ and $J_0\propto u$, $J_1\propto
v$ are the two bosons fields. Now, if we perform the semiclassical
limit and do the appropriate identifications $(4.20)$ yields $(4.19)$
for $\gamma=2$ and from $(4.21)$ we get the following dispersionless
Lax representation
$$
\eqalign{
L=&p+u+vp^{-1}\cr
{\partial L\over\partial t}=&{1\over2}\{(L^2)_{\ge 1},L\}
}\eqno(4.22)
$$
For any $\gamma$ we can use $(4.22)$ as an ansatz to obtain the
dispersionless Lax representation for $(4.19)$ and it reads [5]
$$
\moldura{$\eqalign{
L=&p^{\gamma-1}+u+{v^{\gamma-1}\over(\gamma-1)^2}p^{-(\gamma-1)}\cr
{\partial L\over\partial t}=&{(\gamma-1)\over\gamma}
\left\{\left(L^{\gamma\over\gamma-1}\right)_{\ge1},L\right\}\cr
}$}{6.5}
\eqno(4.23)
$$
In [30] two sets of conserved charges were derived for $(4.19)$ when
$\gamma\ne 2$. So, if $(4.23)$ is really the correct Lax pair it must
somehow provide both sets accordingly to  the algebraic scheme
described in the last section. In fact, since $L$ has singularities in
$p=0$ and $p=\infty$ we can expand $L^{1\over\gamma-1}$ in powers of
$p$ in the two following  ways
$$
\displaylines{
L^{1\over\gamma-1}=p\Biggl\{1+{1\over\gamma-1}\left[up^{-(\gamma-1)}+{v^{\gamma-1}\over(\gamma-1)^2}\,p^{-2(\gamma-1)}\right]+{(2-\gamma)\over2(\gamma-1)^2}\Bigl[\cdots\Bigr]^2+\hfil\Bigg.\cr
\hfill\Bigg.+{(2-\gamma)(3-2\gamma)\over6(\gamma-1)^3}\Bigl[\cdots\Bigr]^3+\cdots\Biggr\}\qquad p\to\infty\qquad(4.24a)\cr
\noalign{\vskip 1.0truecm}
L^{1\over\gamma-1}={vp^{-1}\over(\gamma-1)^{2\over\gamma-1}}\Biggl\{1+{1\over\gamma-1}\left[(\gamma-1)^2v^{-(\gamma-1)}\left(up^{(\gamma-1)}+p^{2(\gamma-1)}\right)\right]+\hfil\Bigg.\cr
\hfill\Bigg.+{(2-\gamma)\over2(\gamma-1)^2}\Bigl[\cdots\Bigr]^2+{(2-\gamma)(3-2\gamma)\over6(\gamma-1)^3}\Bigl[\cdots\Bigr]^3+\cdots\Biggr\}\qquad p\to0\qquad(4.24b)\cr
}
$$
So, the first set of charges follows from\vfill\eject
$$
\moldura{${\overline{\cal H}}_n=\underbrace{\hbox{Tr}\,L^{n+{\gamma-2\over\gamma-1}}}_{p\to\infty}=\int dx\,{\overline H}_n$}{5.2}\quad n=0,1,2,3,\dots
\eqno(4.25)
$$
where the first densities are
$$
\eqalign{
{\overline H}_0=& {(\gamma-2)\over(\gamma-1)}u\cr
{\overline H}_1=& {(2\gamma-3)(\gamma-2)\over(\gamma-1)^2}\left({1\over2!}u^2+
{1\over(\gamma-1)(\gamma-2)}v^{\gamma-1}\right)\cr
{\overline H}_2=& {(3\gamma-4)(2\gamma-3)(\gamma-2)\over(\gamma-1)^3}\left({1\over3!}u^3+
{1\over(\gamma-1)(\gamma-2)}uv^{\gamma-1}\right)\cr
&\vdots\cr
{\overline H}_n=&(n+1)!\hbox{C}^{(n+1)(\gamma-1)-1\over(\gamma-1)}_{n+1}H_{n+1}
}\eqno(4.26)
$$
and
$$
H_n=\sum_{m=0}^{\left[{n\over2}\right]}
\left(-\prod_{k=0}^m{1\over k(\gamma-1)-1}\right)
{u^{n-2m}\over m!(n-2m)!}{v^{m(\gamma-1)}\over(\gamma-1)^m}
\eqno(4.27)
$$
which are the first set of charges obtained in [30]. The second set
follows from
$$
\moldura{${\overline {\widetilde{\cal H}}}_n=\underbrace{\hbox{Tr}\,L^{n+{1\over\gamma-1}}}_{p\to0}=\int dx\,{\overline{\widetilde{H}} }_n$}{5.2}\quad n=0,1,2,3,\dots\eqno(4.28)
$$
and the first densities are
$$
\eqalign{
{\overline {\widetilde H}}_0=& (\gamma-1)^{-{2\over\gamma-1}} v\cr
{\overline {\widetilde H}}_1=& (\gamma-1)^{-{2\over\gamma-1}}
{\gamma\over(\gamma-1)}uv\cr
\overline{{\widetilde H}}_2=& (\gamma-1)^{-{2\over\gamma-1}}
{\gamma(2\gamma-1)\over(\gamma-1)^2}\left({1\over2!}u^2v+
{v^\gamma\over\gamma(\gamma-1)}\right)\cr
&\vdots\cr
{\overline{\widetilde H}}_n=&{n!\over(\gamma-1)^{2\over\gamma-1}}
\hbox{C}^{n(\gamma-1)+1\over(\gamma-1)}_{n}{\widetilde H}_n
}\eqno(4.29)
$$
where
$$
{\widetilde H}_n=
\sum_{m=0}^{\left[{n\over2}\right]}
\left(\prod_{k=0}^m{1\over k(\gamma-1)+1}\right)
{u^{n-2m}\over m!(n-2m)!}{v^{m(\gamma-1)+1}\over(\gamma-1)^m}
\eqno(4.30)
$$
is the second set of charges obtained in [30]

In $(4.23)$  $L^{1\over\gamma-1}$ was expanded in $p=\infty$, a
expansion around $p=0$ provides a second consistent
dispersionless Lax equation
$$
{\partial L\over\partial t}=
\left\{\left(L^{\gamma-2\over\gamma-1}\right)_{\le0},L\right\}
\eqno(4.31)
$$
which yields (with the proper rescaling) the equations
$$
\eqalign{
u_t&=v^{\gamma-3}v_x\cr
v_t&=u_x
}\eqno(4.32)
$$
From the chart in Figure 4 we recognize this equations as the polytropic elastic
media equation. 
\bigskip
\noindent {\bf 4.3 Born-Infeld Equation}
\medskip

With the Lax representation for the polytropic gas, obtained in the last
section, we can get a Lax representation for the Born-Infeld
equation given in the chart of Figure 4
$$
\vbox{\moldura{$\eqalign{u_t=& \left({1\over u^2}+{1\over v^2}\right)u_x-{2u\over v^3}v_x\cr
v_t=&\left({1\over v^2}+{1\over u^2}\right)v_x -{2v\over u^3}u_x\cr}$}{5.4}}
\eqno(4.33)
$$
In (4.33) the Born-Infeld equation is expressed in the so called null
coordinates version [31]. If we perform the transformation
$$
\eqalign{
u&=\phi_x\cr
v&=-{\phi_x\over\sqrt{1+\phi_x\phi_t}}}
\eqno(4.34)
$$
we obtain the Born-Infeld equation written as a second-order equation
in null coordinates
$$
\phi_x^2\phi_{tt}+\phi^2_t\phi_{xx}-(4+2\phi_x\phi_t)\phi_{xt}=0
\eqno(4.35)
$$
A Lax representation for $(4.33)$ can be obtained as follows [6]. In the
first place if we do the change of variables
$$
\eqalign{
{\widetilde u}&=-(u^2+v^2)\cr
{\widetilde v}&=\displaystyle{1\over 2}uv
}\eqno(4.36)
$$
called Verosky transformation [31], we will end up with the equation
$$
\eqalign{
{\widetilde u}_t+{\widetilde u}{\widetilde u}_x+\displaystyle{{\widetilde v}_x\over{\widetilde v}^3}&=0\cr
{\widetilde v}_t+({\widetilde u}{\widetilde v})_x&=0
}\eqno(4.37)
$$
known as the Chaplygin gas. In view of this it would be desirable to
first obtain a Lax description of the Chaplygin gas like equations
$$
\eqalign{
{\widetilde u}_t+{\widetilde u}{\widetilde u}_x+{{\widetilde v}_x\over{\widetilde v}^{\alpha+2}}=&0\,,\quad\alpha\ge1\cr
{\widetilde v}_t+({\widetilde u}{\widetilde v})_x=&0
}\eqno(4.38)
$$
This is indeed possible if we set $\gamma\to-\alpha$, $\alpha\ge1$ in
$(4.23)$, so $(4.38)$ can be obtained from
$$
\eqalign{L=&p^{-(\alpha+1)}+{\widetilde u}+{{\widetilde v}^{-(\alpha+1)}\over(\alpha+1)^2}p^{\alpha+1}\cr
{\partial L\over\partial t}=&{(\alpha+1)\over\alpha}
\left\{\left(L^{\alpha\over\alpha+1}\right)_{\le1},L\right\}\cr
}\eqno(4.39)
$$
where $L^{1\over\alpha+1}$ is expanded around $p=0$ and
$\left(L^{\alpha\over\alpha+1}\right)_{\le1}$ is the polynomial in $p$ that
produces consistent equations, instead of the purely nonnegative
polynomial used in (4.23). For $\alpha=1$ the Lax operator
$$
\moldura{$\eqalign{
L=&p^{-2}-\left({1\over u^2}+{1\over v^2}\right)+{1\over u^2v^2}p^2\cr
{\partial L\over\partial t}=&2
\left\{\left(L^{1\over2}\right)_{\le1},L\right\}\cr
}$}{6.5}
\eqno(4.40)
$$
reproduces $(4.33)$. Again, conserved charges follows from
$$
{\widetilde{\cal
H}}_n=\underbrace{\hbox{Tr}\,L^{n-{1\over2}}}_{p\to\infty}=\int
dx\,{\widetilde{H}}_n\quad n=0,1,2,3,\dots
\eqno(4.41)
$$
and the first Born-Infeld charges are
$$
\eqalign{
{\widetilde H}_0=& -uv\cr
{\widetilde H}_1=& {1\over2}\left({u\over v}+{v\over u}\right)\cr
{\widetilde H}_2=& -{3\over4}
\left({u\over 2v^3}+{3\over uv}+{v\over2u^3}\right)\cr
&\vdots\cr
}\eqno(4.42)
$$
and these are exactly the charges derived in [31]. Another set is
obtained from
$$
{\cal H}_n=\underbrace{\hbox{Tr}\,L^{n+{3\over2}}}_{p\to0}=\int dx\,H_n\quad n=0,1,2,3,\dots\eqno(4.43)
$$
and the first ones are
$$
\eqalign{
H_0=& -{3\over 2}\left({1\over u^2}+{1\over v^2}\right)\cr
H_1=& {15\over8}\left({1\over u^4}+
{11\over6}{1\over u^2v^2}+{1\over v^4}\right)\cr
H_2=& -{35\over 16}\left({1\over u^6}-
{1\over u^4v^2}-{1\over u^2v^4}+{1\over v^6}\right)\cr
&\vdots\cr
}\eqno(4.44)
$$
This is a new set of conserved charges, for the Born-Infeld equation
$(4.33)$, not found previously in [31].

\bigskip
\noindent {\bf 5. Conclusions}
\medskip

We believe, from the results of the Section 4, that the study of
dispersionless systems via a Lax representation is worthwhile. So, the
search for a dispersionless Lax representation for the equations in the
upper part of the chart in Figure 4 is being
pursued. Also, the derivation of the multi-Hamiltonian structures of
these systems, as described in [28-31], is under investigation
following the coadjoint orbit method [32,33]. Another question that
comes to mind  is the dispersive generalization of these
equations. Attempts in this direction can be found in [38].

Some topological equations are also related with the systems discussed
here. For instance, the hyperbolic Monge-Amp\`ere equation
$$
U_{tt}U_{xx}-(U_{tx})^2=-1\eqno(5.1)
$$
may be related with the Born-Infeld equation as follows. If we perform
the change of variables
$$
\eqalign{
a&=U_x\cr 
b&=U_t
}\eqno(5.2)
$$
the Monge-Amp\`ere equation can be written as a first order system
$$
\eqalign{
a_t&=b_x\cr
b_t&={b_x^2-1\over a_x}
}\eqno(5.3)
$$
and this equation can be related to the Chaplygin gas equation $(4.37)$
through the following change of variables
$$
\eqalign{
{\widetilde u}&=-{b_x\over a_x}\cr
{\widetilde v}&=a_x
}\eqno(5.4)
$$
Thus, we can give a Lax description for the hyperbolic Monge-Amp\`ere
equation through the Lax representation derived in Section
4.3. Finally, the Witten-Dijkgraaf-Verlinde-Verlinde (WDVV) equations
$(3.9)$, for $n=3$, with
$$
F(t^1,t^2,t^3)={1\over2}(t^1)^2t^3+{1\over2}t^1(t^2)^2+f(t^2,t^3)\eqno(5.5)
$$ 
where $t^2\equiv x$ and  $t^3\equiv t$, yields
the third order Monge-Amp\`ere equation
$$
f_{ttt}=f_{xxt}^2-f_{xxx}f_{xtt}\eqno(5.6)
$$
This equation is a bi-Hamiltonian system and has a matrix Lax
representation. It is then possible to generate a whole set of
nonlocal charges much like the nonlinear sigma model (details are
given in [7]). It is likely that a dispersionless sort of Lax
representation for (5.6) may exist.
\bigskip
\leftline{\bf Acknowledgments}
\medskip
 
I would like to thank Ashok Das and Celso M. Doria for useful
discussions. This work was supported by CNPq, Brazil.
\vfill\eject
{\leftline{\bf References}
\medskip
\item{1.}{L. D. Faddeev and L. A. Takhtajan, {\it Hamiltonian Methods
in the Theory of Solitons}, Springer-Verlag, 1987.}
\item{2.}{A. Das, {\it Integrable Models}, World Scientific, 1989.}
\item{3.}{P. G. Drazin and R.S. Johnson, {\it Solitons: an
Introduction}, Cambridge University Press, 1989.}
\item{4.}{J. C. Brunelli, {\it Hamiltonian Structures for the
Generalized Dispersionless KdV Hierarchy}, Rev. Math. Phys. {\bf 8}, 1041(1996) (solv-int/961001).}
\item{5.}{J. C. Brunelli and A. Das, {\it A Lax Description for Polytropic Gas Dynamics}, Phys. Lett. {\bf A235}, 597(1997) (solv-int/9706005).}
\item{6.}{J. C. Brunelli and A. Das, {\it A Lax Representation for the Born-Infeld Equation}, Phys. Lett. {\bf B426}, 57(1998) (hep-th/9712081).}
\item{7.}{J. C. Brunelli and A. Das, {\it Non-local Charges and their
Algebra in Topological Field Theory}, Phys. Lett. {\bf B438}, 99(1998) (hep-th/9802070).}
\item{8.}{J. Scott-Russell, {\it Report on Waves}, 14th Meeting of the
British Association for the Advancement of Science, John Murray,
London, 1844, p. 311.}
\item{9.}{D. J. Korteweg and G. de Vries, {\it On the Change of Form
of Long Waves Advancing in a Rectangular Canal, and on a New Type of
Long Stationary Waves}, Philos. Mag. {\bf 39}, 422(1895).}
\item{10.}{E. Fermi, J. Pasta and S. M. Ulam, {\it Studies in
Nonlinear Problems}, Tech. Rep. LA-1940, Los Alamos Sci. Lab. (also in
Collected Papers of Enrico Fermi, Vol. II, 1965, p. 978, Chicago
University Press).}
\item{11.}{N. J. Zabusky and M. D. Kruskal, {\it  Interaction of
``Solitons'' in a Collisionless Plasma and the Recurrence of Initial
States}, Phys. Rev. Lett. {\bf 15}, 240(1965).}
\item{12.}{C. S. Gardner, J. M. Greene, M. D. Kruskal and R. M. Miura,
{\it Method for Solving the Korteweg-de Vries Equation},
Phys. Rev. Lett. {\bf 19}, 1095(1967).}
\item{13.}{P. D. Lax, {\it Integrals of Nonlinear Equations of
Evolution and Solitary Waves}, Comm. Pure Appl. Math. {\bf 21}, 467(1968).}
\item{14.}{C. S. Gardner, {\it Korteweg-de Vries Equation and
Generalizations. IV. The  Korteweg-de Vries Equation as a Hamiltonian
System}, J. Math. Phys. {\bf 11}, 1548(1970).}
\item{15.}{V. E. Zakharov and L. D. Faddeev, {\it  Korteweg-de Vries
Equation: A Completely Integrable Hamiltonian Systems},
Func. Anal. Appl. {\bf 5}, 280(1971).}
\item{16.}{F. Magri, {\it A Simple Model of the Integrable Hamiltonian
Equation}, J. Math. Phys. {\bf 19}, 1156(1978).}
\item{17.}{I. M. Gel'fand and L. A. Dickey, {\it Asymptotic Behaviour
of the Resolvent of Sturm-Liouville Equations and the Algebra of the
Korteweg-de Vries Equation}, Russ. Math. Surveys, {\bf 30}, 77(1975).}
\item{18.}{M. Adler, {\it On a Trace Functional for Formal
Pseudodifferential Operators and the Symplectic Structure for the
Korteweg-de Vries Type Equations}, Invent. Math. {\bf 50}, 403(1979).}
\item{19.}{G. B. Whitham, {\it Linear and Nonlinear Waves}, John Wiley
\& Sons, 1974.}
\item{20.}{V. E. Zakharov, {\it Benney Equations and Quasiclassical
Approximation in the Method of the Inverse Problem},
Funct. Anal. Appl. {\bf 14}, 89(1980).}
\item{21.}{D. Lebedev and Yu. I. Manin, {\it Conservation Laws and Lax
Representation on Benney's Long Wave Equations}, Phys. Lett. {\bf
74A}, 154(1979).}
\item{22.}{V. E. Zakharov, {\it On the Benney's Equations}, Physica
{\bf 3D}, 193(1981).}
\item{23.}{I. Krichever, {\it The Dispersionless Lax Equations and
Topological Minimal Models}, Comm. Math. Phys. {\bf 143}, 415(1992);
{\it The $\tau$-Function of the Universal Whitham Hierarchy, Matrix
Models and Topological Field Theories}, Commun. Pure Appl. Math. {\bf
47}, 437(1994); {\it Whitham Theory for Integrable Systems and
Topological Quantum Field Theories}, in New Symmetry Principles in
Quantum Field Theory, p. 309, ed. J.~Fr\"ohlich et al., Plenum Press,
New York, 1992.}
\item{24.}{B. A. Dubrovin and S. P. Novikov, {\it Hydrodynamics of
Weakly Deformed Soliton Lattices. Differential Geometry and
Hamiltonian Theory}, Russ. Math. Surv. {\bf 44}, 35(1989).}
\item{25.}{B. Dubrovin, {\it Integrable Systems in Topological Field
Theory}, Nucl. Phys. {\bf B379}, 627(1992); {\it Geometry of 2D
Topological Field Theories}, preprint SISSA-89/94/FM, SISSA, Trieste,
1994, hep-th/9407018.}
\item{26.}{E. Witten, {\it On the Structure of the Topological Phase
of Two-Dimensional Gravity}, Nucl. Phys. {\bf B340}, 281(1990).}
\item{27.}{R. Dijkgraaf, H. Verlinde and E. Verlinde, {\it Topological
Strings in $d<1$}, Nucl. Phys. {\bf B352}, 59(1991).}
\item{28.}{Y. Nutku, {\it On a New Class of Completely Integrable
Nonlinear Wave Equations. I. Infinitely Many Conservation Laws},
J. Math. Phys. {\bf 26}, 1237(1985).}
\item{29.}{Y. Nutku, {\it On a New Class of Completely Integrable
Nonlinear Wave Equations. II. Multi-Hamiltonian Structure},
J. Math. Phys. {\bf 28}, 2579(1987).}
\item{30.}{P. J. Olver and Y. Nutku, {\it Hamiltonian Structures for
Systems of Hyperbolic Conservation Laws}, J. Math. Phys. {\bf 29}, 1610(1988).}
\item{31.}{M. Arik, F. Neyzi, Y. Nutku, P. Olver and J. M. Verosky,
{\it Multi-Hamiltonian Structure of the Born-Infeld Equation}, J. Math. Phys. {\bf 30}, 1338(1989).}
\item{32.}{A. A. Kirillov, {\it Elements of the Theory of
Representations}, Springer Verlag, 1976; B. Kostant, {\it Quantization
and Unitary Representation, part I: Prequantization}, Lect. Notes in
Math. {\bf 170}, 87(1970); J. M. Souriau, {\it Structure des Systems
Dynamiques}, Dunod, 1970.}
\item {33.}{B. Kostant, {\it Quantization and Representation Theory},
London Math. Soc. Lect. Notes {\bf 34}, 287(1979); A. G. Reyman,
M. A. Semenov-Tian-Shansky and I. B. Frenkel, {\it Graded Lie Algebras
and Completely Integrable Systems}, Sov. Math. Dokl. {\bf 20},
811(1979);  A. G. Reyman and M. A. Semenov-Tian-Shansky, {\it
Reduction of Hamiltonian Systems, Affine Lie Algebras and Lax
Equations. I and II}, Invent. Math. {\bf 54}, 81(1979) and {\bf
63}, 423(1981); W. W. Symes, {\it Systems of Toda Type, Inverse
Spectral Problems and Representation Theory}, Invent. Math. {\bf 59},
13(1980); D. R. Lebedev and Yu. I. Manin, {\it Gelfand-Dikii
Hamiltonian Operator and Coadjoint Representation of Volterra Group},
Funct. Anal. Appl. {\bf 13}, 268(1980).}
\item{34.}{D. J. Benney, {\it Some Properties of Long Nonlinear
Waves}, Stud. Appl. Math. {\bf 52}, 45(1973).}
\item{35.}{L. J. F. Broer, {\it Approximative Equations for Long Water Waves},
Appl. Sci. Res. {\bf 31}, 377(1975); D. J. Kaup, {\it A Higher-Order
Water-Wave Equation and the Method for Solving It},
Progr. Theor. Phys. {\bf 54}, 396(1975).}
\item{36.}{B. A. Kupershmidt, {\it Mathematics of Dispersive Water
Waves}, Commun. Math. Phys. {\bf 99}, 51(1985).}
\item{37.}{J. C. Brunelli, A. Das and W.-J. Huang ,{\it Gelfand-Dikii
Brackets for Nonstandard Lax Equations}, Mod. Phys. Lett. {\bf 9A},
2147 (1994).}
\item{38.}{B. Enriquez, A. Yu Orlov and V. N. Rubtsov, {\it
Dispersionful Analogues of Benney's Equations and $N$-Wave Systems},
Inverse Problems {\bf 12}, 241(1996).}
\medskip
\end